# ALMA OBSERVATIONS OF LYMAN-α BLOB 1: HALO SUB-STRUCTURE ILLUMINATED FROM WITHIN

J. E. Geach[1], D. Narayanan[2,3], Y. Matsuda[4,5], M. Hayes[6], Ll. Mas-Ribas[7], M. Dijkstra[7], C. C. Steidel[8], S. C. Chapman[9], R. Feldmann[10], A. Avison[11,12], O. Agertz[13], Y. Ao[4], M. Birkinshaw[14], M. N. Bremer[14], D. L. Clements[15], H. Dannerbauer[16,17], D. Farrah[18], C. M. Harrison[19], N. K. Hine[1], M. Kubo[4], M. J. Michałowski[20], Douglas Scott[21], D. J. B. Smith[1], M. Spaans[22], J. M. Simpson[20], A. M. Swinbank[19], Y. Taniguchi[23], E. van Kampen[24], P. van der Werf[25], A. Verma[26], T. Yamada[27]

## ABSTRACT

We present new Atacama Large Millimeter/Submillimeter Array (ALMA) 850μm continuum observations of the original Lyman-α Blob (LAB) in the SSA22 field at $z = 3.1$ (SSA22-LAB01). The ALMA map resolves the previously identified submillimeter source into three components with total flux density $S_{850} = 1.68 \pm 0.06$ mJy, corresponding to a star formation rate of $\sim 150 M_\odot$ yr$^{-1}$. The submillimeter sources are associated with several faint ($m \approx 27$ mag) rest-frame ultraviolet sources identified in *Hubble Space Telescope* Imaging Spectrograph (STIS) clear filter imaging ($\lambda \approx 5850$Å). One of these companions is spectroscopically confirmed with Keck MOSFIRE to lie within 20 projected kpc and 250 km s$^{-1}$ of one of the ALMA components. We postulate that some of these STIS sources represent a population of low-mass star-forming satellites surrounding the central submillimeter sources, potentially contributing to their growth and activity through accretion. Using a high resolution cosmological zoom simulation of a $10^{13} M_\odot$ halo at $z = 3$, including stellar, dust and Lyα radiative transfer, we can model the ALMA+STIS observations and demonstrate that Lyα photons escaping from the central submillimeter sources are expected to resonantly scatter in neutral hydrogen, the majority of which is predicted to be associated with halo substructure. We show how this process gives rise to extended Lyα emission with similar surface brightness and morphology to observed giant LABs.

[1] Centre for Astrophysics Research, University of Hertfordshire, Hatfield, AL10 9AB, UK. j.geach@herts.ac.uk
[2] Dept. of Physics and Astronomy, Haverford College, PA, 19041
[3] Dept. of Astronomy, University of Florida, Gainesville, FL, 32608
[4] NAOJ, Osawa, Mitaka, Tokyo 181-8588, Japan
[5] The Graduate University for Advanced Studies, 2-21-1 Osawa, Mitaka, Tokyo 181-8588, Japan
[6] Stockholm University, Department of Astronomy and Oskar Klein Centre for Cosmoparticle Physics, SE-10691, Stockholm, Sweden
[7] Institute of Theoretical Astrophysics, University of Oslo, P.O. Box 1029 Blindern, NO-0315 Oslo, Norway
[8] California Institute of Technology, 1216 East California Boulevard, MS 249-17, Pasadena, CA 91125
[9] Dept. of Physics and Atmospheric Science, Dalhousie University, Halifax, NS B3H 4R2, Canada
[10] Dept. of Astronomy, University of California Berkeley, CA 94720
[11] UK ALMA Regional Centre Node
[12] Jodrell Bank Centre for Astrophysics, School of Physics and Astronomy, The University of Manchester, Manchester, M13 9PL, UK
[13] Dept. of Physics, University of Surrey, GU2 7XH, Surrey, UK
[14] H. H. Wills Physics Laboratory, University of Bristol, Tyndall Avenue, Bristol, BS8 1TL, UK
[15] Imperial College London, Blackett Laboratory, Prince Consort Road, London SW7 2AZ, UK
[16] Instituto de Astrofísica de Canarias, La Laguna, Tenerife, Spain
[17] Universidad de La Laguna, Astrofísica, La Laguna, Tenerife, Spain
[18] Dept. of Physics, Virginia Tech, Blacksburg, VA 24061
[19] Centre for Extragalactic Astronomy, Department of Physics, Durham University, South Road, Durham, DH1 3LE, UK
[20] Institute for Astronomy, University of Edinburgh, Royal Observatory, Blackford Hill, Edinburgh, EH9 3HJ, UK
[21] Dept. of Physics & Astronomy, University of British Columbia, Vancouver, BC, V6T 1Z1, Canada
[22] Kapteyn Astronomical Institute, University of Groningen, PO Box 800, 9700 AV Groningen, Netherlands
[23] The Open University of Japan, Wakaba, Mihama-ku, Chiba, 261-8586, Japan
[24] ESO, Karl-Schwarzschild-Str. 2, D-85748 Garching, Germany
[25] Leiden Observatory, Leiden University, P.O. Box 9513, NL-2300 RA Leiden, The Netherlands
[26] Oxford Astrophysics, Department of Physics, University of Oxford, Keble Road, Oxford, OX1 3RH, UK
[27] Tohoku University, Sendai, Miyagi 980-8578, Japan

## 1. INTRODUCTION

SSA22-LAB01 ($z = 3.1$, Steidel et al. 2000) is the most thoroughly studied (Chapman et al. 2004; Bower et al. 2004; Geach et al. 2007, 2014; Matsuda et al. 2007, Weijmans et al. 2010; Hayes et al. 2011; Beck et al. 2016) giant (100 kpc in projected extent) Lyman-α 'Blob' (LAB). LABs are intriguing objects: the origin of the extended Lyα emission could be due to gravitational cooling radiation, with pristine hydrogen at $T \sim 10^{4-5}$ K cooling primarily via collisionally excited Lyα as it flows into young galaxies (Katz et al. 1996; Haiman et al 2000; Fardal et al. 2001). Alternatively, galactic winds, photoionization, fluorescence or scattering processes have been proposed (Taniguchi & Shioya 2000; Geach et al. 2009, 2014; Hine et al. 2016; Alexander et al. 2016). In any scenario the picture is one of an extended circumgalactic medium (CGM) that is rich in cool gas (be it clumpy or smoothly distributed), and so LABs reveal astrophysics associated with the environment on scales comparable to the virial radius of the massive dark matter halos they trace. Nevertheless, the process (or processes) giving rise to the extended line emission remains in question.

Cen & Zheng (2013) predicted that giant LABs should contain far-infrared sources close to the gravitational centre of the halo. The presence of a central submillimetre-bright galaxy within SSA22-LAB01 has been in debate (Chapman et al. 2004; Matsuda et al. 2007; Yang et al. 2012), but was recently confirmed with the solid detection of an unresolved $S_{850} = 4.6 \pm 1.1$ mJy submillimetre source using the SCUBA-2 instrument on the 15-m James Clerk Maxwell Telescope (JCMT) (Geach et al. 2014). Here we present new high-resolution Band 7 continuum ($\lambda_{\rm obs} = 850$μm) observations of SSA22-LAB01 with the Atacama Large Millimeter/Submillimeter Array (ALMA). These observations resolve the rest-frame 210-μm emission at $z = 3.1$, close to the peak of the thermal dust emission and therefore a good probe of the cold and dense interstellar medium (ISM). Throughout we assume a cosmology with $\Omega_\Lambda = 0.72$, $\Omega_m = 0.28$ and $h = 0.697 = H_0/100$ km s$^{-1}$ Mpc$^{-1}$. At $z = 3$ $1''$ subtends approximately 8 projected kiloparsecs (pkpc).

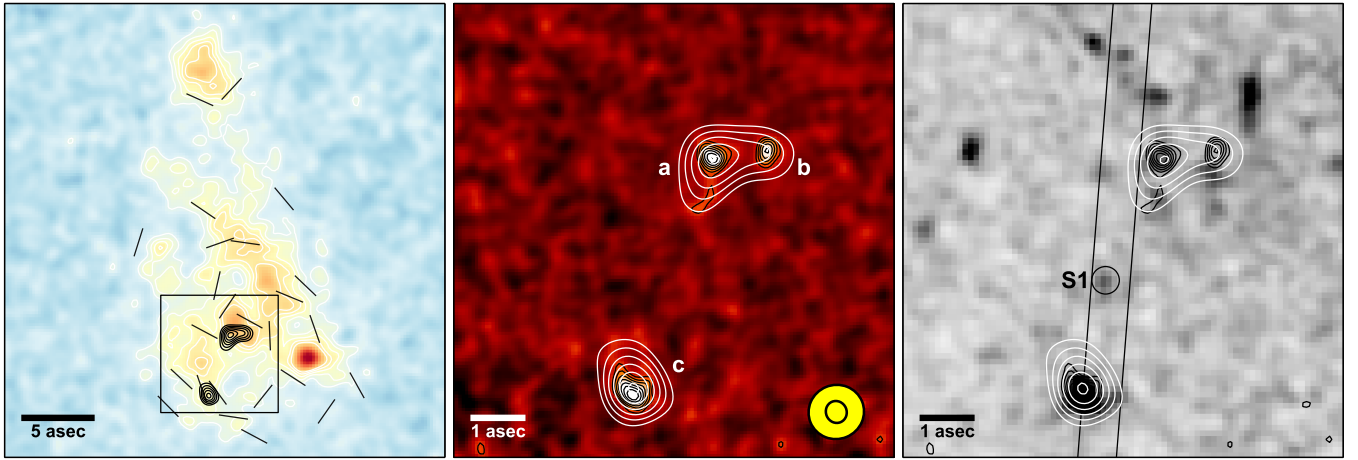

**Figure 1.** Observations of SSA22-LAB01. (left) MUSE continuum-subtracted line image showing the Lyα emission averaged over 4976–5000Å, with contours at levels of 0.5, 0.7 and $1\times10^{-19}$ erg s$^{-1}$ cm$^{-2}$ Å$^{-1}$. Black lines show Lyα polarization (Hayes et al. 2011) and black contours show the $1''$ tapered ALMA 850μm emission at $>3\sigma$ significance; (center) zoom-in showing the ALMA map with the highest resolution image as background with the full resolution and tapered 850μm contours overlaid (starting at $3\sigma$, increasing in steps of $1\sigma$), yellow ellipses show the FWHM of the full resolution and tapered synthesized beams; (right) *HST* STIS optical image of the same region as the central panel, indicating the orientation of the Keck MOSFIRE slit (§2.4) and the position of a faint companion source we label 'S1'. We overplot the same contours as the central panel to illustrate how the submillimeter sources associate with emission in the STIS map.

## 2. OBSERVATIONS

### 2.1. *Atacama Large Millimeter/Submillimeter Array*

SSA22-LAB01 was observed in two projects (2013.1.00922.S [PI Geach] and 2013.1.00704S [PI Matsuda]) with a similar configuration in Band 7: the full 7.5 GHz of bandwidth centred at 347.59 GHz was used to measure the continuum emission at approximately the same frequency as the SCUBA-2 detection of the target. A total of 36–40 12 metre antennas were used in the observations, with baselines spanning 21–918 m, thus capable of recovering emission on scales of up to $5''$, with a maximum resolution of $0.4''$. Antenna $T_{sys}$ temperatures were approximately 100–150 K and the mean precipitable water vapour column was 0.249 mm. Observations of the phase calibrator source J2206-0031 confirm consistent flux scaling between the two projects in the calibration phase. All calibration was performed using the CASA software and the visibilities from each project concatenated into a single measurement set for which the total integration time is 2860 seconds.

The dirty image reveals the two main components of submillimetre emission, and we use this information to supply circular masks (each $2''$ in radius) at $22^h17^m26.0^s$, $+00°12'36.3''$ and $22^h17^m26.1^s$, $+00°12'32.4''$ (J2000) to the CASA clean task. We use a natural weighting of the visibilities and clean down to a threshold of $70\mu$Jy within 1000 iterations. We adopt multi-scale cleaning, allowing model Gaussian components of width $0''$ (delta function), $0.5''$, $1''$, $2''$ and $3''$. The clean is run twice: at full resolution and with an outer taper of $1''$. To both maps we 'feather' the single dish SCUBA-2 (regridded) map to improve the recovery of the total flux of the source and potentially improve sensitivity to extended emission. The depth of the map is $40 \mu$Jy beam$^{-1}$, increasing to $90 \mu$Jy beam$^{-1}$ after tapering. Final maps are corrected for primary beam attenuation for analysis. The synthesized beams are $0.40'' \times 0.38''$ (PA $23°$) and $1.01'' \times 0.98''$ (PA $112°$) in the feathered full resolution and $1''$ tapered maps respectively.

To measure the integrated flux densities and sizes we threshold the maps at $3\sigma$, where $\sigma$ is the root mean square noise measured around the sources and sum the flux enclosed by each $3\sigma$ contour. In the full resolution map there are three distinct components: a, b and c. Integrated flux density uncertainties are estimated from the root mean squared value, scaled to the number of beams subtended by the $3\sigma$ contour. The sizes of each source in the tapered map are measured as the maximum width of the $3\sigma$ contour, with an uncertainty $\delta\theta = 0.6 \times \theta_{beam}/SNR$, where $\theta_{beam}$ is the full width half maximum of the beam and SNR = 3.

Note that there is a discrepancy between the single dish flux and the total flux measured in the ALMA map: $\Delta S_{850} = 2.9 \pm 1.1$ mJy. Although this is not significant at the $3\sigma$ level, we should discuss this potential 'missing' flux. First, it is important to note that single dish flux densities in the submillimeter can be statistically boosted; Geach et al. (2016) measured the flux boosting as a function of signal-to-noise ratio in the SCUBA-2 Cosmology Legacy Survey (where the SCUBA-2 detection of this target originates), finding an average boosting of approximately 30% for SSA22-LAB01. The array configuration allows us to recover emission on scales of up to $5''$. It is unlikely that there is a 850μm emission component on scales larger than this (40 pkpc at this redshift), but given the sensitivity of our observations, it implies that, if real, the missing submillimetre emission is extended on scales of at least $3''$ (25 pkpc). One possibility is that the emission is spread over a large number of faint clumps around the central sources – an idea we explore later in the paper. Deeper observations with a slightly more compact array configuration would be beneficial to address this.

### 2.2. *Multi-Unit Spectroscopic Explorer*

Integral field spectroscopy was performed with the Very Large Telescope Multi Unit Spectroscopic Explorer (MUSE) instrument. The observations were carried out under clear or photometric conditions, between November 2014 and September 2015. We used the extended blue setting, resulting in a minimum wavelength of 4650Å. Each integration was 1500 seconds, and the field was rotated by 90° between each exposure in order to reduce fixed pattern noise from the integral field units (IFUs) and residual flat-field errors. Data were reduced with the MUSE pipeline (version 1.0.5), following standard procedures for bias subtraction, dark current removal, flat-fielding, and basic calibration of the individual integrations. Each individual spectrum was fully post-processed to output an image of the field of view, data cubes, and pixel tables, which we examined individually. We computed shifts between the individual exposures using standard centering tools in IRAF. Finally, we used the MUSE_EXP_COMBINE task to drizzle and stack all the individually reduced pixtables.

### 2.3. *Hubble Space Telescope Imaging Spectrograph*

The *Hubble Space Telescope* Imaging Spectrograph (STIS) observations (project 9174, Chapman et al. 2004) used the 50CCD clear filter which provides a wide response in the optical, with effective central wavelength 5850Å and full width half maximum throughput 4410Å over a $52'' \times 52''$ field-of-view. Six exposures of SSA22-LAB01 were taken in 'LOW-SKY' time for a total of 7020 seconds of integration. *HST* Legacy Archive calibrated images were combined into a single deep co-add with $0.1''$ pixels using the DRIZZLEPAC software (version 2) from the Space Telescope Science Institute. The $5\sigma$ sensitivity limit is 27.6 mag (Chapman et al. 2004).

### 2.4. *Keck Multi-Object Spectrometer For Infra-Red Exploration*

*K*-band (1.92–2.30μm) spectroscopy of SSA22-LAB01 was obtained on a single slit of a multislit mask observed during science verification of the Multi-Object Spectrometer For Infra-Red Exploration (MOSFIRE, Mclean et al. 2012) on the W. M. Keck Observatory Keck 1 10m telescope. The slit width was $0.7''$, resulting in a spectral resolving power of $R \sim 3600$. The total integration time was 5040 seconds, in a sequence of 28 180s exposures using an ABAB nod pattern with a nod amplitude of $3.0''$. The seeing during the observation was estimated to be $0.35''$ full width half maximum. The data were reduced by the MOSFIRE Data Reduction Pipeline (Steidel et al. 2014).

## 3. ANALYSIS & INTERPRETATION

In the following sections we analyse and interpret the observations, characterising the ALMA sources in the context of the LAB environment. We then use a high resolution cosmological hydrodynamic simulation and radiative transfer to develop a hypothesis about the role of the central submillimeter sources in contributing to the extended Lyα emission, paying particular attention to creating mock observations that can be directly compared to the data, and determining whther systems resembling SSA22-LAB01 are expected in current models of galaxy formation.

### 3.1. *Observations*

At full ($0.4''$) resolution we resolve three main components that we label 'a', 'b' and 'c' within the SSA22-LAB01 (Figure 1). Component 'a' is the brightest submillimetre source, located close to the geometric centre of the LAB and is $1''$ from component 'b'; in the tapered map these effectively blend into a single source, and we consider a+b jointly hereafter. Component 'c' is located $4.5''$ to the south-east and is associated with strong stellar continuum emission, with a precise redshift measured from [OIII] and Hβ lines, $z = 3.1000 \pm 0.0003$, placing it within the Lyα nebula (Kubo et al. 2015). Component a+b is also coincident with stellar continuum emission of complex, clumpy morphology – it is not clear whether this is a single bound system and we are simply detecting bright clumps of submillimeter and optical emission, or we are witnessing the coalescence of several independent low mass systems.

**Table 1**
Properties of ALMA sources in SSA22-LAB01.

| Source | R.A. (h m s) | Dec. (o ′ ″) | $S_{850}$ (mJy) |
|---|---|---|---|
| SSA22-LAB01 ALMA a+b | 22 17 26.01 | +00 12 36.5 | $0.95 \pm 0.04$ |
| SSA22-LAB01 ALMA c | 22 17 26.11 | +00 12 32.2 | $0.73 \pm 0.05$ |

The total flux density of signal above the $3\sigma$ level in the tapered map is $S_{850} = 1.68 \pm 0.06$ mJy; Table 1 gives the positions and flux density of the two main components a+b and c. Assuming cool dust emission traces the dense ISM (Scoville et al. 2015), the corresponding total molecular gas mass is $M_{\rm gas} \approx 4 \times 10^{10}$ $M_\odot$ and the average molecular gas surface density is $\Sigma_{\rm gas} \approx 140\,M_\odot$ pc$^{-2}$. If $\Sigma_{\rm SFR}$ scales linearly with $\Sigma_{\rm gas}$ (Wong & Blitz 2002) we estimate that the submillimetre sources are forming stars at rates of order $100\,M_\odot$ yr$^{-1}$. We have poor constraints on the shape of the spectral energy distribution of these sources, but we can estimate their integrated total infrared luminosities by assuming an appropriate range of templates describing the thermal dust emission of active galaxies. We adopt the Dale & Helou (2002) set of templates, characterised by a parameter $\alpha$, describing the power law index for the distribution of dust mass over heating by the interstellar radiation field ($U$): $dM(U) \propto U^{-\alpha}dU$, with most galaxies falling within the range $\alpha = 1$–2.5; we conservatively consider this full range when estimating the total infrared luminosities of the sources. Redshifting to $z = 3.1$ and normalising to the observed 850μm flux density, we estimate total infrared (8–1000μm) luminosities in the range $L_{\rm IR} \approx (0.2 - 1.9) \times 10^{12} L_\odot$ and $L_{\rm IR} \approx (0.2 - 1.5) \times 10^{12} L_\odot$ for for sources a+b and c respectively. The median infrared luminosity considering the full range of templates is $L_{\rm IR} \approx 6.0 \pm 0.3 \times 10^{11} L_\odot$ and $L_{\rm IR} \approx 4.6 \pm 0.3 \times 10^{11} L_\odot$ for the two components respectively, where the error bar reflects the measurement uncertainty on the integrated ALMA flux. Assuming a star formation rate calibration following Kennicutt & Evans (2012) these average luminosities correspond to SFR $\approx 90 M_\odot$ yr$^{-1}$ and SFR $\approx 70 M_\odot$ yr$^{-1}$ respectively, consistent with the ISM mass scaling value derived above, although the SFRs could be a factor of three higher if the galaxies are better represented by the more 'active' templates of the Dale & Helou library.

The emergent luminosity of Lyα emission associated with obscured star formation is $L_{\rm Ly\alpha} = 0.05 f_{\rm esc,Ly\alpha} L_{\rm IR}$ (Djikstra & Westra 2010), under the assumption of Case B conditions and where $f_{\rm esc,Ly\alpha}$ is the escape fraction (of Lyα photons) from the galaxy. This corresponds to $L_{\rm Ly\alpha} \sim f_{\rm esc,Ly\alpha} 10^{44}$ erg s$^{-1}$ per ALMA source, similar to the total Lyα luminosity of SSA22-LAB01. From the models of Leitherer et al. (1999), the ionizing flux ($\nu = 200$–912Å) scales like $L_{200-912} \approx {\rm SFR} \times f_{\rm esc,UV} \times 10^{43}$ erg s$^{-1}$ (for 100 Myr of continuous star formation, Solar metallicity and Salpeter IMF). Thus, for star formation rates of order $100 M_\odot$ yr$^{-1}$ and modest high escape fractions, the ALMA sources appear to be viable 'power sources' for extended Ly$\alpha$ emission through photoionization. The next question follows naturally: what is the nature of the environment around these central star-forming sources?

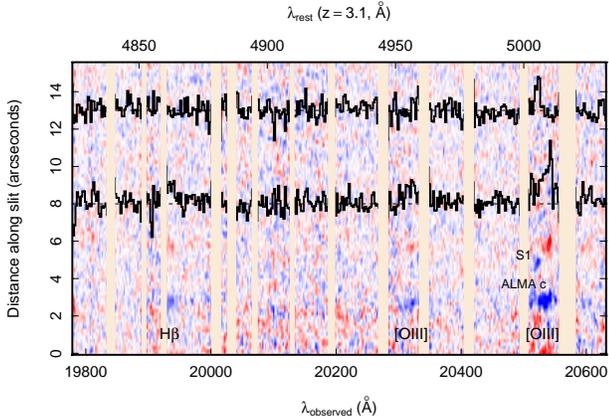

**Figure 2.** Near-infrared spectrum of ALMA source c and companion obtained with Keck MOSFIRE. The background image shows the two dimensional spectrum, slightly smoothed for clarity. Strong atmospheric emission lines are masked. The nod pattern employed results in a positive (red) and negative (blue) signal in the co-added spectrum offset by 3″. The one dimensional spectra for each source reveal H$\beta$ and [O III] emission lines, with the 5007Å [O III] line separated by 250 km s$^{-1}$ between ALMA source 'c' and S1, which are separated by an angular distance of 2.1″ (17 pkpc).

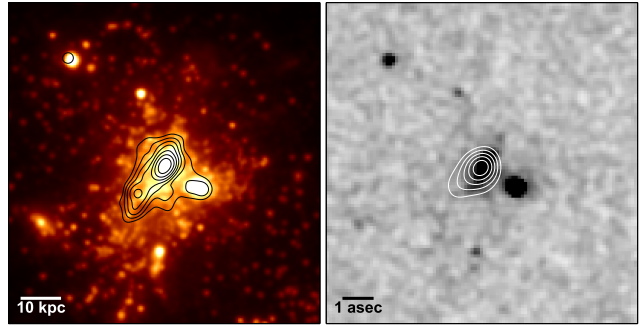

**Figure 3.** Predicted ultraviolet and submillimetre emission of a $M_h \approx 10^{13} M_\odot$ halo at $z \approx 3$ from a hydrodynamic simulation with realistic radiative transfer (§3.2). (left) predicted emission in the *HST* STIS 50CCD band with zero noise. Black contours show the predicted 850$\mu$m emission, logarithmically spaced at 0.25 dex, starting at 0.1$\mu$Jy. Both the STIS and submillimeter image has been slightly smoothed for clarity; (right) shows the same emission, but as would be observed with an identical set-up to the real observations. Background scaling of the mock *HST* STIS image and 1″ tapered ALMA contours are identical to Figure 1 (we only show the tapered contours for clarity here). Only the brightest handful of neighbouring STIS sources are detectable, with flux densities of $f_\lambda \gtrsim 2 \times 10^{-20}$ erg s$^{-1}$ cm$^{-2}$ Å$^{-1}$. Formally, we detect five STIS sources above a $3\sigma$ detection threshold, which is similar to the observations after considering contamination from projections (§3.3).

Like other giant LABs (e.g. Prescott et al. 2012), SSA22-LAB01 is thought to trace a massive dark matter halo of order $M_h \approx 10^{13} M_\odot$; this is close to the mass where cold flows are not expected to survive inside the virial radius at this epoch (Dekel et al. 2009). Recently, moving mesh hydrodynamic simulations suggest cold streams are assimilated into the hot halo gas inside the virial radius even down to halo masses of $M_{halo} \approx 10^{11.5} M_\odot$ (Nelson et al. 2013), suggesting that cold flows are not sufficient to explain the extended Ly$\alpha$ luminosity of giant LABs. Following arguments put forward by Hayes et al. (2011), Geach et al. (2014) and Beck et al. (2016), we postulate that extended line emission in SSA22-LAB01 could arise from Ly$\alpha$ photons generated by central star formation in the submillimetre sources, escaping and resonantly scattering in neutral gas associated with a large ensemble of lower-mass galaxies in the same potential well (e.g. Cen & Zheng 2013; Francis et al. 2013). This is consistent with recent spectropolarimetry observations of SSA22-LAB01 that favour a scenario in which Ly$\alpha$ photons undergo long flights from central sources before scattering in H I in the CGM (Hayes et al. 2011; Beck et al. 2016, although cf. Trebitsch et al. 2016). In Figure 1 we show the polarization vectors of Hayes et al. (2011) that describe a circular pattern roughly centered on the submillimeter sources.

The STIS imaging reveals many faint ($f_\lambda \approx 10^{-19}$ erg s$^{-1}$ cm$^{-2}$ Å$^{-1}$) clumps of UV emission surrounding the submillimetre sources (Figure 1, Chapman et al. 2004; Matsuda et al. 2007, Uchimoto et al. 2008, see also Prescott et al. 2012). There is no spectroscopically confirmed redshift for source a+b, and, unlike source c, it is not unambiguously associated with a STIS counterpart. Instead, there are a number of STIS clumps at (and around) the position of a+b, and the optical colors of these are broadly consistent with a $z = 3.1$ redshift (Kubo et al. 2015). In addition, there is a possible H I absorption feature in the Ly$\alpha$ spectrum at this location (Weijmans et al. 2010; Geach et al. 2014); thus, source a+b is unlikely to be a chance projection of a submillimetre source at a different redshift. In the event that this ALMA source is not associated with the LAB, then obviously the main impact on our analysis and interpretation is that there will be approximately a factor two fewer Ly$\alpha$ and UV photons generated by central star formation.

If some (a fraction are likely to be chance projections) of the other faint STIS sources represent low mass satellites of the ALMA sources, then the cold gas associated with them could be responsible for scattering Ly$\alpha$ photons emerging from the central active galaxies. A key test is actually confirming 'membership' of the STIS sources, and to this end we have obtained near-infrared spectroscopic confirmation of one of these faint companion sources 'S1' (Figure 2), detecting the [O III] line at $z = 3.0968$. S1 is 2.1″ (17 pkpc) from ALMA source 'c' with a line-of-sight velocity offset of just 250 km s$^{-1}$. Although not a complete survey by any means, this observation does support the view that the central ALMA sources are accompanied by a population of low-mass satellites. With this empirical evidence in hand, we now consider whether the same observations are predicted by galaxy formation and evolution models. In particular, we run a zoomed cosmological hydrodynamical simulation coupled with radiative transfer schemes to produce mock observations that can be directly compared to the data. The main objective is to test whether systems similar to SSA22-LAB01 arise in current galaxy formation schemes.

### 3.2. *Simulations*

We simulate the formation of a galaxy in a massive ($M_h \approx 10^{13} M_\odot$) halo (by $z = 2$) utilising a cosmological zoom technique with the hydrodynamics code GIZMO (Hopkins et al. 2014). We employ a pressure-entropy formulation of smoothed particle hydrodynamics, which conserves momentum, energy, angular momentum, and entropy. We first run a coarse large (144 Mpc$^3$) dark matter-only simulation to identify a halo of interest. This simulation is run at a relatively low mass resolution $M_{dm} = 8.4 \times 10^8 M_\odot$. At $z = 2$, we select a massive ($2.03 \times 10^{13} M_\odot$) halo, and re-simulate the evolution of this halo at significantly higher resolution (Feldmann et al. 2016), including baryons with particle mass $m_{bar} = 2.7 \times 10^5 M_\odot$. The initial conditions are generated with the MUSIC code (Hahn et al. 2011) and the minimum baryonic, star and dark matter force softening lengths are, respectively, 9, 21 and 142 proper parsecs at $z = 2$.

Gas cools using a cooling curve that includes both atomic and molecular line emission, with the neutral ISM broken into atomic and molecular components following an analytic algorithm describing the molecular fraction in a cloud based on the gas surface density and metallicity (Krumholz, McKee & Tumlinson 2009). Star formation occurs in molecular gas and follows a volumetric relation $\dot{\rho}_\star = \rho_{mol}/t_{ff}$, where $\rho_{mol}$ is the volume density of molecular gas and $t_{ff}$ is the freefall time. Star formation is restricted to gas that is locally self-gravitating following the algorithms of Hopkins, Narayanan & Murray (2013). Once stars have formed, they impact the ISM via a number of feedback channels. In particular, we include models for momentum deposition by stellar radiation and supernovae, photoheating of HII regions, and stellar winds (see Hopkins et al. 2014 for details). GIZMO tracks 11 metal species (H, He, C, N, O, Ne, Mg, Si, S, Ca, and Fe) with yields from Type 1a and Type II supernova following Iwamoto et al. (1999) and Woosley & Weaver (1995).

To model the ALMA and STIS observations we employ the dust radiative transfer code POWDERDAY which computes the stellar spectral energy distributions (SEDs) of the star particles (constrained by their ages and metallicities) and propagates this radiation through the dusty ISM. The gas particles are projected onto an adaptive grid with an octree-like memory structure (using YT, Turk et al. 2011), and the dust mass is assumed to be a constant fraction (40%) of the total metal mass in a given cell (Dwek et al. 1998; Watson 2011). Photons are emitted in a Monte Carlo fashion and absorbed, re-radiated, and scattered by dust; this process is iterated upon until the dust temperature has converged. The stellar SEDs are calculated at run time for each stellar particle utilising FSPS (Conroy & Gunn 2010). We employ the Padova isochrones and assume a Kroupa IMF. The dust radiative transfer employs HYPERION (Robitaille et al. 2011). We assume a Weingartner & Draine (2001) dust size distribution, with $R_V = 3.15$. To create mock observations we consider the observed frame 850$\mu$m and optical emission predicted by POWDERDAY and HYPERION projected onto a celestial grid at the position of SSA22-LAB01.

To simulate the ALMA observations we use the CASA task SIMALMA, setting the simulation parameters to match the real observations (§2.1), including the same array configuration. The resulting maps have nearly identical synthesized beam and 1$\sigma$ noise as the data. For the STIS imaging we convolve the predicted observed frame UV-through-optical emission with the 50CCD throughput curve to generate single images projected along the same view angles as the 850-$\mu$m images. These images are then convolved with the STIS point spread function and gridded to an identical pixel scale as the observations. Finally, we create a noise image by removing >1.5$\sigma$ features from the real STIS image and replace those pixels with values randomly drawn from the remainder of the pixels in the image. The noise-free, PSF-convolved flux image is then added to the noise image to create the mock observation.

Finally, we consider Ly$\alpha$ radiative transfer of centrally generated photons through the simulation volume. We use the code TLAC (Gronke et al. 2014, 2015) with the default core skipping parameter $x = 3$. The central 400 pkpc of simu-lation volume is interpolated onto a regular grid, and we treat two central star-forming galaxies (SFRs 189 $M_\odot$ yr$^{-1}$ and 115 $M_\odot$ yr$^{-1}$) as sources of Ly$\alpha$ photons. To compute the Ly$\alpha$ luminosity emitted by the cells, we use the relation between star formation rate and Ly$\alpha$ luminosity (applicable to solar metallicity and a Kroupa IMF)

$$L_{\text{Ly}\alpha}/\text{erg s}^{-1} = 1.87 \times 10^{42} \times (\text{SFR}/M_\odot \text{ yr}^{-1}) . \quad (1)$$

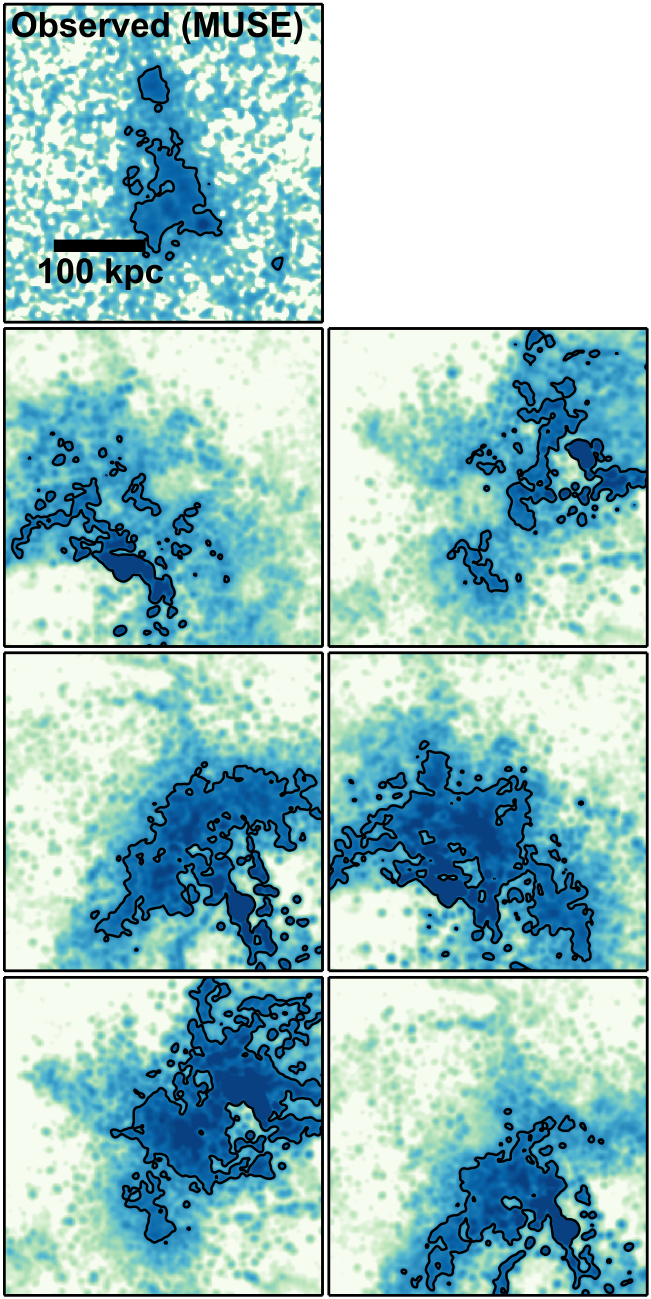

**Figure 4.** Predicted surface brightness profiles of extended Ly$\alpha$ emission compared to observations. The first panel shows the MUSE observation of SSA22-LAB01 with an isophotal contour at $\mu_{Ly\alpha} = 2.2 \times 10^{-18}$ erg s$^{-1}$ cm$^{-2}$ arcsec$^{-2}$ used to originally classify LABs (Matsuda et al. 2004). Additional panels show, at the same scale, the Ly$\alpha$ surface brightness maps predicted by our simulation (no noise has been added), smoothed with a Gaussian beam of FWHM 1″ to mimic ground-based seeing. These maps are generated by projecting the simulation volume along each of the six faces of the cube. The central star-forming galaxies (analogous to the ALMA sources) are at the center of the cube, and so it is clear that large scale scattering in the halo substructure can form LABs that are highly asymmetric about the 'power' sources, and produce emission line halos with a range of morphology. Note also the fainter extended emission predicted by the model on scales larger then the single isophote shown.

The photons are emitted from a random point inside the corresponding cell with an emission frequency drawn from a distribution depending on the thermal velocity of the gas in the cell (Gronke et al. 2014, 2015) using the 'peeling' algorithm (Yusef et al. 1984; Zheng et al. 2002; Laursen 2010). We follow the prescriptions of Laursen (2010) for the Small Magellanic Cloud (Pei et al. 1992; Weingartner & Draine 2001; Gnedin et al. 2008) when calculating the dust content in the grid, and a full description of this method can be found in that work. We compute an effective dust density $n_d$ at every cell as

$$n_d = (n_{\text{HI}} + f_{\text{ion}} n_{\text{HII}})(Z/\langle Z_{\text{SMC}}\rangle), \quad (2)$$

where $n_{\text{HI}}$ and $n_{\text{HII}}$ are the neutral and ionised hydrogen densities respectively at every cell. $Z$ and $\langle Z_{\text{SMC}}\rangle$ denote the metallicity at every cell and the SMC average respectively, and $f_{\text{ion}}$ accounts for the dust content in the ionised gas (Laursen 2010). We assume that $f_{\text{ion}}$ is 1% and note that $\langle Z_{\text{SMC}}\rangle$ is 0.6 dex below the solar value.

### 3.3. *Comparing data and simulations*

Figure 3 shows the synthetic ALMA and STIS observations. We do not expect perfect fidelity with the observations, but we can note some important similarities. As in the observation, the simulated halo contains two main star-forming galaxies with SFRs 189 and 115 $M_\odot$ yr$^{-1}$ respectively; these are well within the range of SFRs we estimate in §3.1. Both galaxies are submillimeter emitters, but only one has a flux density high enough to be detected in our mock ALMA map, with $S_{850} \approx 0.5$ mJy, within a factor of two of the observations. In the simulation, both galaxies are bright STIS sources, whereas this is only true of one of three observed ALMA components. The two central galaxies are surrounded by several STIS-detectable clumps (within ∼5″ or about 40 pkpc) that are of similar flux to those observed. Formally, we detect five sources at the $>3\sigma$ level in the synthetic STIS image[28], with flux densities $f_\lambda \approx (0.2$–$4) \times 10^{-19}$ erg s$^{-1}$ cm$^{-2}$ Å$^{-1}$. This is compared to eight sources with $f_\lambda \approx (0.2$–$2) \times 10^{-19}$ erg s$^{-1}$ cm$^{-2}$ Å$^{-1}$ over the same area in the real image using identical detection criteria (Figure 1). In the real data some of these sources could be chance projections, and we quantify this contamination by running the detection on the wider STIS image to measure the average surface density of sources in the same flux range. In a random 5″ radius aperture we would expect 4±1 sources by chance, so the predicted abundance of STIS sources in the simulation is in good agreement with the observations.

What are these faint neighbouring sources? In the simulation, these STIS clumps are the brightest of a population of star-forming sub-halos (that are apparent in the left panel of Figure 3) that 'swarm' the central star-forming galaxies galaxy, bombarding it over a ∼1 Gyr period in a phase of hierarchical growth that accompanies the accretion of recycled gas previously heated by stellar feedback at an earlier epoch (Narayanan et al. 2015). So the simulation can broadly reproduce the ALMA and STIS observations, and we can interpret this as a demonstration that, as far as can be determined by the data, the halo population observed in SSA22-LAB01 is consistent with the prediction of this particular model of galaxy evolution for a halo of equivalent mass. What of the extended Ly$\alpha$ emission?

---

[28] We detect sources using SEXTRACTOR version 2.19.5 (Bertin & Arnouts 1996) with a detection threshold of 3 contiguous pixels above 3× the r.m.s. in the local background. No filtering is applied, and we use a back size of 32 pixels.

Considering the two main star-forming galaxies sources as Ly$\alpha$ emitters (with Ly$\alpha$ luminosities scaled from their SFRs, Equation 1), we find that 58% of centrally generated photons escape the simulation box, and 16% of those scatter at least once in H I in the CGM. The majority (77%) of the H I mass in the central 400 pkpc of the simulation volume is tied to subhalos, and so it follows that the majority of the Ly$\alpha$ scattering occurs in and around satellites, which can be treated as cold clumps within the hot halo gas. Figure 4 shows the projected Ly$\alpha$ surface brightness maps of the simulated halo compared to the latest integral field observations from MUSE. Resonant scattering in the satellite population gives rise to extended Ly$\alpha$ emission with surface brightness distributions similar to observed LABs ($\mu_{\text{Ly}\alpha} \approx 10^{-18}$ erg s$^{-1}$ cm$^{-2}$ arcsec$^{-2}$). The observed Ly$\alpha$ surface brightness and morphology is highly orientation dependent Ly$\alpha$ sources – the emission need not be symmetric about the central sources that are the source of the Ly$\alpha$ photons. This is simiular to the situation in real LABs, for example, the next largest LAB in SSA22, SSA22-LAB02 contains an x-ray luminous AGN (Geach et al. 2009; Alexander et al. 2016) that is significantly offset from the peak and geometric center of the LAB. This has important implications for correctly identifying potential luminous galaxy counterparts to LABs, the apparent absence of which in some systems has previously been argued in favour of cold mode accretion (e.g. Nilsson et al. 2006, but see also Prescott et al. 2015).

## 4. SUMMARY

It is uncontroversial that the intergalactic medium is rich in cool baryons at $z \approx 3$, and recent observations have demonstrated how gas in the cosmic web can be detected in Ly$\alpha$ emission on scales much larger than 100 pkpc via illumination by nearby quasars (Cantalupo et al. 2014; Martin et al. 2015). Deep stacking experiments have also demonstrated that diffuse, extended Ly$\alpha$ emission appears to be a generic feature within 100 pkpc of star-forming galaxies at high-redshift, potentially linked to scattering in a clumpy CGM (Steidel et al. 2011). However it has not been clear how cold gas is actually distributed within dark matter haloes and how it relates to the growth of central galaxies.

Our new ALMA observations have resolved two active galaxies at the heart of SSA22-LAB01, with a total star formation rate of order 150 $M_\odot$ yr$^{-1}$, and there is evidence to suggest that these are surrounded by a retinue of lower mass satellites, the brightest of which are detectable in deep *HST* STIS imaging. Our self-consistent hydrodynamic simulation and mock observations lead us to postulate that cold gas associated with the halo substructure – much of which is predicted to be associated with the satellites – acts as a scattering medium for Ly$\alpha$ photons escaping from the central star-forming galaxies, and demonstrate that this can give rise to large Ly$\alpha$ nebulae similar to observed LABs. Although resonant scattering is unlikely to be the only contributor to the extended Ly$\alpha$ emission, we argue that it may be a dominant process. We suggest that deep, high resolution Lyα observations offer a route to mapping the distribution and kinematics of the satellite populations – or, more generally, sub-structure – of distant massive haloes.

## ACKNOWLEDGEMENTS

We thank the referee for a constructive report that has improved this work. J.E.G. is supported by a Royal Society University Research Fellowship. Partial support for D.N. was provided by NSF AST-1442650, NASA HST AR-13906.001 and a Cottrell College Science Award. M.H. acknowledges the support of the Swedish Research Council, Vetenskapsradet and the Swedish National Space Board (SNSB), and is Fellow of the Knut and Alice Wallenberg Foundation. H.D. acknowledges financial support from the Spanish Ministry of Economy and Competitiveness (MINECO) under the 2014 Ramón y Cajal program MINECO RYC-2014-15686. A.V. is supported by a Research Fellowship from the Leverhulme Trust. The authors acknowledge excellent support of the UK ALMA Regional Centre Node. This letter makes use of the following ALMA data: ADS/JAO.ALMA#2013.1.00704S, ADS/JAO.ALMA#2013.1.00922.S. ALMA is a partnership of ESO (representing its member states), NSF (USA) and NINS (Japan), together with NRC (Canada) and NSC and ASIAA (Taiwan), in cooperation with the Republic of Chile. The Joint ALMA Observatory is operated by ESO, AUI/NRAO and NAOJ. This work was supported by the ALMA Japan Research Grant of NAOJ Chile Observatory, NAOJ-ALMA-0086.